# Predictions of conventional and microscopic triaxial cranking models for light nuclei

P. Gulshani
NUTECH Services, Mississauga, Ontario, Canada L5L 5N1, parviz.gulshani@outlook.com

The conventional cranking model for uniaxial rotation is frequently used to study rotational features in deformed nuclei. However, the model uses a constant angular velocity. To investigate the effect of a dynamic angular velocity, a quantal microscopic time-reversal and $D_2$ invariant cranking model for triaxial rotation (MSCRM3) including residual correction terms is derived from a unitary transformation of the nuclear Schrodinger equation and using Hartree-Fock approach. Except for the angular velocity and residual terms, MSCRM3 and the conventional cranking model for triaxial rotation (CCRM3) Schrodinger equations are identical in form, and are solved iteratively in a similar manner. The article identifies the differences in the rotational features predicted by CCRM3 and MSCRM3 for $^{20}$Ne, $^{24}$Mg, and $^{28}$Si using a self-consistent deformed harmonic-oscillator potential. The rotational features studied are: rotational relaxation of the intrinsic system, stability of the rotational states, various rotation modes, nuclear shapes, their transitions, and band termination. MSCRM3 predicts the observed reduced energy-level spacing in $^{20}$Ne between *J*=6 and 8 and attributes its occurrence to quenching of a wobbly rotation. The remaining discrepancy between the observed and MSCRM3-predicted excitation energies for $^{20}$Ne is removed by including the spin-orbit interaction and the residuals of the square of the angular momentum and interaction. CCRM3 does not predict the three dimensional phenomena predicted by MSCRM3 (such as the reduced energy-level spacing in $^{20}$Ne and the rotational-band termination at *J*=12 in prolate $^{28}$Si and triaxial $^{24}$Mg, etc. arising from the angular velocity). We, therefore, conclude that CCRM3 is effectively a uniaxial rotation model. Therefore, using CCRM3 or its uniaxial version, one would miss capturing three-dimensional rotation phenomena.

## 1. Introduction

The conventional cranking model for triaxial collective rotation (CCRM3) is described by the Schrodinger equation:

$$\hat{H}_{cr}\left|\Phi_{cr}\right\rangle \equiv \left(\hat{H}_o - \vec{\Omega}\cdot\hat{\vec{J}}\right)\left|\Phi_{cr}\right\rangle \equiv E_{cr}\left|\Phi_{cr}\right\rangle \qquad (1)$$

where $\hat{H}_o$ is the nuclear Hamiltonian, $\hat{\vec{J}}$ is the angular-momentum-operator vector, and $\vec{\Omega}$ is the angular-velocity vector. The model is frequently used to predict the uniaxial-rotation properties of deformed nuclei [1-20]. However, it is recognized [2,3,21] that the model is semi-classical and phenomenological. It uses a constant adjustable angular velocity $\vec{\Omega}$ parameter. Therefore, Eq. (1) is time-reversal and $D_2$ and signature non-invariant, and ignores the interaction between $\vec{\Omega}$ and $\hat{\vec{J}}$. There have been many revealing investigations [3,6,21-25] to derive the model from first principles and reveal the implicit model assumptions and approximations using various methods. The major difficulty in deriving a tractable microscopic quantal cranking model has been the strong coupling between the rotation and intrinsic motions.

In Section 2 of this article, we derive a microscopic, quantal, self-consistent, time-reversal and $D_2$-invariant cranking model for triaxial rotation (MSCRM3) containing no free parameters. MSCRM3 is derived from a unitary-rotation transformation of the nuclear Schrodinger equation to an intrinsic rotor equation, and applying Hartree-Fock (HF) method to the rotor equation. The intrinsic-rotation coupling in the transformed equation is completely accounted for using a rigid-flow prescription for the rotation angles in the unitary transformation, and requiring the angles to be canonically conjugate to the angular momentum. The method used in this article contrasts with that used in [3,6,21-25], which were mostly limited to rotation in two dimensions and used a significant number of major approximations and assumptions such as large deformation, large expectation of an angular momentum component, expansion in powers of the angular momentum, angular momentum projection, and generator co-ordinates method using various simplifying approximations such as axial symmetry, expansion in density matrices and limiting the analysis to single-particle density matrix, etc.. In particular, reference [23], which uses an approach somewhat similar to that in this article, assumes a rotation in two dimensions and uses an angular-momentum-projected well-deformed independent-particle HF ground state possessing a large value of the expectation of the square of the angular momentum. It obtains a Hamiltonian that depends on the angular momentum $\hat{J}$ and $\hat{J}^2$ operators where the $\hat{J}$ term involves a product of $\hat{J}$ and the particle momenta. This $\hat{J}$ term represents the coupling between the rotational and intrinsic motions. Reference [23] then minimizes the expectation of $\hat{H}$ in a particle-hole HF approximation up to second order in $\hat{J}^2$ to obtain an expression for the effective moment of inertia similar to that of Thouless-Valatin.

In Section 3, we compare the predictions of MSCRM3 with those of CCRM3 for light nuclei using the self-consistent deformed harmonic oscillator potential, which has been used in [2-5,7,8,10-19] to provide physically transparent results with the least number of adjustable parameters. It is reasonable to use this potential in this article since our objective is to compare the results predicted by MSCRM3 with those by CCRM3 but not with measured data. We use spin-orbit and residuals of nuclear interaction and the square of the angular momentum only for the case of $^{20}$Ne to compare the predicted and measured excitation energies.

Section 4 summarizes MSCRM3 derivation, predictions, and the differences between MSCRM3 and CCRM3 predictions, and presents conclusions.

## 2. Derivation of MSCRM3

We derive MSCRM3 and residual corrections to it by starting with the nuclear Schrodinger equation:

$$\hat{H}_o\left|\Phi\right\rangle \equiv \left(\sum_{n=1}^{A}\frac{\hat{p}_n^2}{2M}+\hat{V}\right)\left|\Phi\right\rangle = E\left|\Phi\right\rangle \qquad (2)$$

where the interaction $\hat{V}$ is rotationally invariant and $A$ is the mass number. Now we perform a dynamic (i.e., nucleon-generated) collective rotation, through the angle $\vec{\theta}$, of the wavefunction $\left|\Phi\right\rangle$ to obtain the wavefunction $\left|\Psi\right\rangle$ as follows:

$$\left|\Phi\right\rangle = e^{-i\vec{\theta}\cdot\hat{\vec{J}}/\hbar}\cdot\left|\Psi\right\rangle \equiv e^{-Z}\cdot\left|\Psi\right\rangle \qquad (3)$$

where the function $\left|\Psi\right\rangle$ is determined below, and the three rotation angles $\vec{\theta}$ and hence the three components of the angular-momentum operator $\hat{\vec{J}}$ describe rotations about the three space-fixed frame axes[1]. Since $\vec{\theta}$ is chosen to be a vector function of the particle coordinates, the rotation operator $e^{-Z}$ in Eq. (3) is rotationally invariant, and hence the wavefunctions $\left|\Phi\right\rangle$ and $\left|\Psi\right\rangle$ are eigenfunctions of the angular momentum and are isotropic. Inserting Eq. (3) for $\left|\Phi\right\rangle$ into Eq. (2), we obtain the Schrodinger equation:

$$\begin{aligned}\hat{H}\left|\Psi\right\rangle &\equiv e^{Z}\cdot\hat{H}_o\cdot e^{-Z}\cdot\left|\Psi\right\rangle \\ &= \left(\frac{1}{2M}\sum_{n=1}^{A}e^{Z}\cdot\hat{p}_n^2\cdot e^{-Z}+\hat{V}\right)\left|\Psi\right\rangle = E\left|\Psi\right\rangle\end{aligned} \qquad (4)$$

We note that, in Eq. (4), $\vec{\theta}$ and $\hat{p}_n^2$ do not commute. The transformed Hamiltonian $\hat{H}$ in Eq. (4) is observed to be rotationally invariant since:

$$\begin{aligned}\left[\hat{\vec{J}},e^{Z}\cdot\hat{p}_n^2\cdot e^{-Z}\right] &= e^{Z}\left[\hat{\vec{J}},\hat{p}_n^2\right]e^{-Z} = 0 \\ \left[\hat{\vec{J}},\hat{V}\right] &= 0\end{aligned} \qquad (5)$$

The invariance in Eq. (5) is one of the major advantages of the rotation operator $e^{Z}$ over other forms of the rotation operators used in the literature.

We evaluate $e^{z}\cdot\hat{p}_n^2\cdot e^{-z}$ in Eq. (4) using the following expansion in powers of $Z$:

$$e^{Z}\cdot\hat{p}_n^2\cdot e^{-Z} = \hat{p}_n^2+\left[Z,\hat{p}_n^2\right]+\frac{1}{2!}\left[Z,\left[Z,\hat{p}_n^2\right]\right]+.... \qquad (6)$$

Using the rotational invariance of $Z$ and $\hat{p}_n^2$ in Eq. (5), we can easily evaluate the terms in Eq. (6) and show that the terms in powers of $Z$ higher than two vanish since $\vec{\theta}$ is a function of only the nucleon co-ordinates. Inserting these results into Eq. (4), we obtain:

$$\begin{aligned}\hat{H} = \hat{H}_o &+ \frac{i\hbar}{2M}\cdot\sum_{n=1}^{A}\sum_{k=1}^{3}\nabla_n^2\theta_k\cdot\hat{J}_k \\ &- \frac{1}{M}\cdot\sum_{n=1}^{A}\sum_{k=1}^{3}\left(\vec{\nabla}_n\theta_k\hat{\vec{p}}_n\right)\cdot\hat{J}_k \\ &+ \frac{1}{2M}\cdot\sum_{n=1}^{A}\sum_{k=1}^{3}\vec{\nabla}_n\theta_k\cdot\vec{\nabla}_n\theta_l\cdot\hat{J}_k\cdot\hat{J}_l\end{aligned} \qquad (7)$$

[1] We use the space-fixed-frame-axis components of the rotation-angle $\vec{\theta}$ and angular-momentum operator vectors because they simplify the analysis and yields a transformed $\hat{H}$ that is a function symmetric in the square of the angular momentum components along the space-fixed-frame axes, whereas the Euler angles about the body-fixed-frame axes yield an $\hat{H}$ that is complicated asymmetric function of the angular momentum components.

To express the third term in Eq. (7) in terms of $\vec{J}$, we use the following rigid-flow velocity-field prescription for the rotation angles $\vec{\theta}$:

$$\frac{\partial \theta_l}{\partial x_{nj}} = -\sum_{k\neq l\neq j=1}^{3} {}_l\chi_{jk}\, x_{nk}, \quad {}_l\chi_{jk} = -{}_l\chi_{kj} \qquad (8)$$

$$\left(l, j, k = 1,2,3\right)$$

where each of the three $3\times 3$ matrices ${}_l\chi$ is real and anti-symmetric. The prescription in Eq. (8) renders $\hat{H}$ in Eq. (4) dependent on only $\hat{J}_k^2$ [2]. The non-zero elements of the matrices ${}_l\chi$ are determined by choosing $\theta_k$ and $\hat{J}_k$ to be a canonically conjugate pair, which together with Eq. (5) yields:

$${}_l\chi_{jk} = \frac{1}{\hat{\mathcal{J}}_l}, \quad \hat{\mathcal{J}}_l \equiv \sum_{n=1}^{A}\left(x_{nj}^2 + x_{nk}^2\right), \qquad (9)$$

$$j,k,l = 1,2,3 \text{ and in cyclic order}$$

$M\hat{\mathcal{J}}_l$ is the $l^{th}$ principal-axis component of the rigid-flow moment of inertia tensor in the space-fixed frame[3]. We observe that, for the choice in Eqs. (8) and (9), we obtain the result:

$$\left[\theta_k, \hat{J}_l\right] = i\hbar\frac{Q_{kl}}{\hat{\mathcal{J}}_k} \approx 0\left(k\neq l\right) \qquad (10)$$

since $Q_{lk} \equiv \sum_{n=1}^{A} x_{nl}\cdot x_{nk}$ (the $kl^{th}$ component of the quadrupole moment tensor) for $k \neq l$ is very small compared to $\hat{\mathcal{J}}_k$. (However, there is no need to know the action of $\hat{J}_l$ on $\theta_k$ for $k\neq l$ in any derivations of the equations in this article.) Inserting Eqs. (8) and (9), into Eq. (7), we obtain:

$$\hat{H} \equiv \hat{H}_o - \sum_{k=1}^{3}\frac{1}{2M\hat{\mathcal{J}}_k}\cdot\hat{J}_k^2 - \sum_{k\neq l=1}^{3}\frac{Q_{lk}}{2M\hat{\mathcal{J}}_l\hat{\mathcal{J}}_k}\cdot\hat{J}_l\cdot\hat{J}_k \qquad (11)$$

Inserting Eq. (11) into Eq. (4), we obtain the following Schrodinger equation for a triaxial rotation of a microscopic quantum triaxial rotor:

$$\hat{H}\left|\Psi\right\rangle \equiv \left(\hat{H}_o - \sum_{k=1}^{3}\frac{\hat{J}_k^2}{2M\hat{\mathcal{J}}_k} - \sum_{k\neq l=1}^{3}\frac{Q_{lk}}{2M\hat{\mathcal{J}}_l\hat{\mathcal{J}}_k}\cdot\hat{J}_l\cdot\hat{J}_k\right)\left|\Psi\right\rangle = E\left|\Psi\right\rangle \qquad (12)$$

Eq. (12) shows that $\left|\Psi\right\rangle$ is not an eigenfunction of $\hat{H}_o$ but rather it is an eigenstate of the reduced Hamiltonian $\hat{H}$.

One can easily show that:

$$\left[\theta_k, \hat{H}\right] = \frac{-i\hbar}{\hat{\mathcal{J}}_k}\left(\frac{\hat{Q}_{kl}}{\hat{\mathcal{J}}_l}\hat{J}_l + \frac{\hat{Q}_{kk'}}{\hat{\mathcal{J}}_{k'}}\hat{J}_{k'}\right) \approx 0, \qquad (13)$$

$$\left(k\neq l\neq k' = 1,2,3\right)$$

since $\hat{Q}_{lk}$ for $k\neq l$ is very small compared to $\hat{\mathcal{J}}_l$. The result in Eq. (13) indicates that $\hat{H}$ in Eq. (12) is nearly independent of $\hat{J}_l$ and hence $\hat{H}$ is optimally or ideally intrinsic. The Hamiltonian $\hat{H}$ is also time-reversal and rotationally invariant since the rotation operator $e^Z$ in Eq. (3) is rotationally invariant.

The inverse of the rigid-flow moment of inertia $\hat{\mathcal{J}}_l$ is a many-body operator. To render Eq. (12) solvable, we replace $\hat{\mathcal{J}}_l$ in Eq. (12) by its expectation value:

$$\mathcal{J}_l \equiv \left\langle\Psi\right|\hat{\mathcal{J}}_l\left|\Psi\right\rangle \qquad (14)$$

This is a reasonably good approximation since $\hat{\mathcal{J}}_l$ is a relatively large number and varies little and gradually with the angular momentum quantum number *J*. Eq. (12) then becomes:

---

[2] For any other angle prescription, higher order terms in $\hat{J}_k$ coupled to other intrinsic (such as shear) operators would appear in $\hat{H}$ in Eq. (4). Therefore, the collective rigid-flow velocity-field prescription in Eq. (8) accounts in an optimum way for the interaction between the angular momentum (i.e., collective rotation) and other types of operators (intrinsic motion), which is manifested in the appearance of the rigid-flow kinematic moment of inertia $\hat{\mathcal{J}}_k$ in Eqs. (9) and (11).

[3] The rigid-flow prescription for $\theta_k$ in Eqs. (8) and (9) is a collective analogue of the Birbrair's single-particle $\theta_n$ [26]. $\theta_n$ has continuous second-order mixed derivatives in any spatial region that excludes the rotation *x* axis along which $\theta_n$ is singular. Whereas $\theta_k$ has discontinuous second-order mixed derivatives. The difference in the discontinuity between $\vec{\nabla}_n\theta_k$ and $\vec{\nabla}_n\theta_n$ arises because of the many-body nature of $\left(\hat{\mathcal{J}}_k\right)^{-1}$ in Eq. (9). However, the discontinuity in $\vec{\nabla}_n\theta_k$ is small because the expectation of $\hat{\mathcal{J}}_k$ is a large number. On the other hand, for a rigid-body motion, with a constant angular velocity, $\vec{\nabla}_n\theta_{rig}$ is inherently discontinuous. Nevertheless, $\theta_{rig}$ is commonly used in the analysis of a rigid-body motion [26,27] with impunity because the second mixed derivatives of $\theta_{rig}$ do not appear in any of the analysis equations. Similarly, the discontinuity in $\vec{\nabla}_n\theta_k$ is of no consequence in the derivation of MSCRM3 because the second-order mixed derivatives of $\theta_k$ do not appear anywhere in the derivation of the equations.

$$\hat{H}\left|\Psi\right\rangle \equiv \left( \hat{H}_o - \sum_{k=1}^{3} \frac{\hat{J}_k^2}{2M\mathcal{J}_k} - \sum_{k\neq l=1}^{3} \frac{\hat{Q}_{lk}}{2M\mathcal{J}_l\,\mathcal{J}_k} \cdot \hat{J}_l \cdot \hat{J}_k \right)\left|\Psi\right\rangle = E\left|\Psi\right\rangle \tag{15}$$

We now apply to Eq. (15) HF variational method to obtain:

$$\hat{H}\left|\Psi\right\rangle = \left[ \hat{H}_{crHF} + \hat{V}_{res} - \sum_{k=1}^{3} \frac{\left(\hat{J}_k^2\right)_{res}}{2M\mathcal{J}_k} \right]\left|\Psi\right\rangle = E\left|\Psi\right\rangle \tag{16}$$

where $\hat{H}_{crHF}$ is the HF mean-field independent-particle part of $\hat{H}$ and satisfies the Schrodinger equation:

$$\begin{aligned} &\hat{H}_{crHF}\left|\Psi_{crHF}\right\rangle \\ &\equiv \left( \hat{H}_{oHF} - \vec{\Omega}\cdot\vec{\hat{J}} - \sum_{l\neq k=1}^{3} \frac{Q_{lk}}{\mathcal{J}_k} \cdot \Omega_l \cdot \hat{J}_k - \frac{1}{2}\sum_{l\neq k=1}^{2} \Omega_k \cdot \Omega_l M \hat{Q}_{lk} \right)\left|\Psi_{crHF}\right\rangle \\ &\equiv E_{crHF}\left|\Psi_{crHF}\right\rangle \end{aligned} \tag{17}$$

where $Q_{lk} \equiv \left\langle \Psi_{crHF}\left|\hat{Q}_{lk}\right|\Psi_{crHF}\right\rangle \equiv \left\langle \hat{Q}_{lk}\right\rangle$. $\hat{H}_{oHF}$ in Eq. (17) is the HF mean-field independent-particle part of $\hat{H}_o$. $\vec{\Omega}\cdot\vec{\hat{J}}$ in Eq. (17) is the single-particle direct HF mean-field part of $\sum_{k=1}^{3}\frac{\hat{J}_k^2}{2M\mathcal{J}_k}$, and $\vec{\Omega}$ is the angular-velocity vector with the magnitude $\Omega$ and orientation $(\theta,\phi)$, and is defined by its three Cartesian components:

$$\begin{aligned} \vec{\Omega} &\equiv \left(\Omega_1, \Omega_2, \Omega_3\right) \\ &\equiv \Omega\left(\sin\theta\cos\phi, \sin\theta\sin\phi, \cos\theta\right) \\ &= \left( \frac{\left\langle\hat{J}_1\right\rangle}{M\mathcal{J}_1}, \frac{\left\langle\hat{J}_2\right\rangle}{M\mathcal{J}_2}, \frac{\left\langle\hat{J}_3\right\rangle}{M\mathcal{J}_3} \right) \end{aligned} \tag{18}$$

$\left(\hat{J}_i^2\right)_{res}$ in Eq. (16) is the residual of the square of the angular-momentum operator given by:

$$\left(\hat{J}_i^2\right)_{res} \equiv \hat{J}_i^2 - 2\left\langle \Psi_{crHF}\left|\hat{J}_i\right|\Psi_{crHF}\right\rangle \cdot \hat{J}_i \tag{19}$$

For the residual of nuclear interaction $\hat{V}_{res}$ in Eq. (16), we may use the commonly-used separable effective quadrupole-quadrupole (long range) interaction.
In Eq. (16) we do not show the residual of the three-body operator $-\sum_{l\neq k=1}^{2}\frac{\hat{Q}_{lk}}{2M\mathcal{J}_l\mathcal{J}_k}\cdot\hat{J}_l\cdot\hat{J}_k$ in Eq. (15) because it is negligibly small, complicated, and is not used in this article.

Eq. (17) shows that the cranking model wavefunction $\left|\Psi_{crHF}\right\rangle$ is an eigenstate of the cranking Hamiltonian $\hat{H}_{crHF}$, which is the HF mean-field part of the intrinsic rotor Hamiltonian $\hat{H}$ in Eq. (15). Since for open-shell nuclei, $\left|\Psi_{crHF}\right\rangle$ describes an anisotropic nucleon spatial distribution, the residual terms in Eq. (16) transforms $\left|\Psi_{crHF}\right\rangle$ into $\left|\Psi\right\rangle$, which is isotropic and an eigenstate of the angular momentum. Generally, we may expect these residuals to be relatively small to render the HF wavefunction $\left|\Psi_{crHF}\right\rangle$ useful and meaningful. Therefore, we approximate $\left|\Psi\right\rangle$ by $\left|\Psi_{crHF}\right\rangle$ and hence CCRM3 ignores the residual terms in Eq. (16) and the small quadrupole terms in Eq (17). When these terms are neglected, Eq. (17) becomes:

$$\begin{aligned} \hat{H}_{crHF}\left|\Psi_{crHF}\right\rangle &= \left(\hat{H}_{oHF} - \vec{\Omega}\cdot\vec{\hat{J}}\right)\left|\Psi_{crHF}\right\rangle \\ &= E_{crHF}\left|\Psi_{crHF}\right\rangle \end{aligned} \tag{20}$$

Since $\vec{\Omega}$ and $\vec{\hat{J}}$ change sign on time reversal, $\hat{H}_{crHF}$ is time-reversal, $D_2$, and signature invariant. This means that the MSCRM3 wavefunction is a linear superposition of either even or odd angular-momentum eigenstates, in contrast to a CCRM3 wavefunction that is a superposition of even and odd angular-momentum eigenstates.

MSCRM3 Eq. (20) is identical in form to CCRM3 Eq. (1) except for the microscopic angular-velocity $\vec{\Omega}$. As is done for CCRM3 Eq. (1), we now present the remaining equations needed to solve iteratively Eq. (20) using a self-consistent deformed harmonic oscillator potential $M\sum_{k=1}^{3}\omega_k^2 x_k^2 \big/ 2$. Taking the expectation of $\hat{H}_{crHF}$ in Eq. (20), we obtain the intrinsic energy:

$$\begin{aligned} E_{int} \equiv E_{crHF} &= \left\langle \hat{H}_{oHF} - \vec{\Omega}\cdot\vec{\hat{J}} \right\rangle = \hbar\sum_{k=1}^{3}\alpha_k\Sigma_k \\ \Sigma_k &\equiv \sum_{n_k=1}^{n_{kf}}\left(n_k+1\right) \end{aligned} \tag{21}$$

and the rotational-band excited-state ($E_J$) and excitation ($\Delta E_J$) energies are then determined from:

$$\begin{aligned} E_J \equiv \left\langle \hat{H}_o \right\rangle &\equiv \left\langle \hat{H}_{crHF} + \vec{\Omega}\cdot\vec{\hat{J}} \right\rangle = E_{\text{int}} + \vec{\Omega}\cdot\left\langle \vec{\hat{J}} \right\rangle \\ \Delta E_J &\equiv E_J - E_{J=0} \end{aligned} \tag{22}$$

where $\alpha_k$ are the normal-mode frequencies and are determined from the equations of motion for $\hat{H}_{crHF}$ in Eq. (20) to be the three roots of the characteristic equation [15]:

$$\alpha^6 - A_o\alpha^4 + B_o\alpha^2 - C_o = 0 \tag{23}$$

where $A_o$, $B_o$, and $C_o$ are functions of $\omega_k^2$.

The expectations of $\hat{J}_k$ and $\sum_{n=1}^{A} x_{nk}^2$ are then given by (using Feynman's theorem):

$$\begin{aligned} \left\langle \hat{J}_k \right\rangle &= -\frac{\partial E_{\text{int}}}{\partial \Omega_k'} = -\hbar \sum_{l=1}^{3} \frac{\partial \alpha_l}{\partial \Omega_k'} \Sigma_l \\ \frac{M}{2}\left\langle \sum_{n=1}^{A} x_{nk}^2 \right\rangle &\equiv \frac{M}{2}\left\langle x_k^2 \right\rangle = \frac{\partial E_{\text{int}}}{\partial \omega_k^2} = \hbar \sum_{l=1}^{3} \frac{\partial \alpha_l}{\partial \omega_k^2} \Sigma_l \end{aligned} \tag{24}$$

Eqs. (24) must also satisfy the constant nuclear-volume ($\omega_1\omega_2\omega_3 = \omega_o^3$) and the density-potential shapes self-consistency conditions. This implies, in the HF mean-field sense, a minimization of $E_{\text{int}}$ with respect to the oscillator frequencies $\omega_k^2$, yielding the result:

$$\omega_1^2\left\langle x^2 \right\rangle = \omega_2^2\left\langle y^2 \right\rangle = \omega_3^2\left\langle z^2 \right\rangle \tag{25}$$

where $\omega_o$ is the isotropic nuclear harmonic-oscillator frequency given by $\hbar\omega_o \equiv P_o A^{-1/3}$.

In Eq. (20), $\left|\ \right\rangle \equiv \left|\Psi_{crHF}\right\rangle$ is required to yield the angular momentum constraint (as in CCRM3):

$$\begin{aligned} \left\langle \hat{J}^2 \right\rangle &\equiv \left\langle \hat{J}_1 \right\rangle^2 + \left\langle \hat{J}_2 \right\rangle^2 + \left\langle \hat{J}_3 \right\rangle^2 = \hbar^2 J(J+1) \\ &\textit{for a triaxial rotation} \\ \left\langle \hat{J}^2 \right\rangle &\equiv \left\langle \hat{J}_1 \right\rangle^2 + \left\langle \hat{J}_2 \right\rangle^2 + \left\langle \hat{J}_3 \right\rangle^2 = \hbar^2 J^2 \\ &\textit{for a principal-axis rotation} \end{aligned} \tag{26}$$

where $J$ is the desired (or externally imposed) angular-momentum quantum number. However, unlike that in CCRM3, the $J_{cal}$ value calculated from $\hbar J_{cal} = \left\langle \hat{J}^2 \right\rangle^{1/2}$ for principal-axis rotation or from $\hbar\sqrt{J_{cal}\left(J_{cal}+1\right)} \equiv \left\langle \hat{J}^2 \right\rangle^{1/2}$ for triaxial rotation is generally different from $J$ because of the interaction between $\left\langle \hat{J}_k \right\rangle$ and $\Omega_k$ generated by Eq. (18) (or equivalently by (28) given below) and the resulting rotational relaxation of the intrinsic system, causing $J_{cal}$ determined from Eq. (26) to be either smaller or larger than the desired or imposed $J$ value.

The orientation angles $(\theta,\phi)$ and the magnitude $\Omega$ of $\vec{\Omega}$ are readily determined from Eq. (18), and are given by:

$$\tan\phi = \frac{\left\langle \hat{J}_2 \right\rangle \mathscr{J}_1}{\left\langle \hat{J}_1 \right\rangle \mathscr{J}_2}, \quad \tan\theta = \frac{\left\langle \hat{J}_1 \right\rangle \mathscr{J}_3}{\left\langle \hat{J}_3 \right\rangle \mathscr{J}_1}\sqrt{1+\tan^2\phi} \tag{27}$$

$$\begin{aligned} \Omega^2 &= \Omega_1^2 + \Omega_2^2 + \Omega_3^2 \\ &= \frac{\left\langle \hat{J}_1 \right\rangle^2}{M^2\mathscr{J}_1^2} + \frac{\left\langle \hat{J}_2 \right\rangle^2}{M^2\mathscr{J}_2^2} + \frac{\left\langle \hat{J}_3 \right\rangle^2}{M^2\mathscr{J}_3^2} \\ &= \left(T_1 + T_2 + T_3\right)/3 \end{aligned} \tag{28}$$

where $\mathscr{J}_k$ is given in Eqs. (9) and (14), and $\left\langle \hat{J}^2 \right\rangle$ is defined in Eq. (26), and:

$$\begin{aligned} T_j \equiv \frac{\left\langle \hat{J}^2 \right\rangle}{M^2\mathscr{J}_j^2} &+ \left(\frac{1}{M^2\mathscr{J}_k^2} - \frac{1}{M^2\mathscr{J}_j^2}\right)\left\langle \hat{J}_k \right\rangle^2 \\ &+ \left(\frac{1}{M^2\mathscr{J}_l^2} - \frac{1}{M^2\mathscr{J}_j^2}\right)\left\langle \hat{J}_l \right\rangle^2 \\ &(j,k,l = 1,2,3 \text{ are in cyclic order}) \end{aligned}$$

## 3. MSCRM3 and CCRM3 predictions and their differences

MSCRM3 Eqs. (14), (20), (21), and (23)-(28) are solved iteratively for each desired $J$ value and for ground-state nucleon configurations of the nuclei $^{20}Ne$, $^{24}Mg$, and $^{28}Si$. An identical set of CCRM3 equations are similarly solved except that in Eqs. (27) $\mathscr{J}_k$ is unity and Eq. (28) is not used but instead $\Omega$ is adjusted to give the desired $J$ value. We use a linear-mixing method, where the input and output values of the variables in each iteration step are linearly mixed and used as input in the next iteration step, to obtain steady converged solutions. From Eq. (22), the values of $E_J$ and $\Delta E_J$ are computed for each desired value of $J$ and the initial values $\phi_o, \theta_o$ of $\phi$, $\theta$. The initial values $\phi_o$ and $\theta_o$ determine the nature of the subsequent rotation.

Since the governing CCRM3 and MSCRM3 equations are similar in form, their predictions are similar. But there are also cases that show significant differences. We now present the latter cases and compare them with the CCRM3 predictions reported in [7,8,10,12-19].

CCRM3 always predicts uniform rotation, i.e., the $\vec{\Omega}$ and $\left\langle \vec{J} \right\rangle$ are parallel vectors. In contrast, MSCRM3 predicts wobbly rotation (compare Eq (27) for the two models).

The CCRM3 and MSCRM3 rotational states are inherently unstable because the self-consistency condition in Eq. (25) causes $E_J$ in Eq. (22) to increase and $E_{\text{int}}$ in Eq. (21) to decrease with $J$. This generates a destabilizing positive-feedback between $\Omega_k$ and $\langle \hat{J}_k \rangle$ (consistently with the first of Eq. (24)) so that a decrease in $\langle \hat{J}_k \rangle$ requires $\Omega_k$ to decrease, which in turn causes $\langle \hat{J}_k \rangle$ to decrease further until it is reduced to zero, unless this free fall is prevented by some constraint such as that in Eq. (26)[4].

For the constraint in Eq. (26), a MSCRM3 and CCRM3 rotational state does not totally collapse rotationally (i.e., all three components $\langle \hat{J}_k \rangle$ simultaneously decrease to zero). For this reason and as seen in Fig. 1, $\langle \hat{J}_3 \rangle$ normally decreases to zero, resulting in a planar uniform rotation in the *x-y* plane in CCRM3 and a planar wobbly rotation in the *x-y* plane in MSCRM3. Then, any small decrease in, for example, $\langle \hat{J}_1 \rangle$ results in a monotonic decrease in $\langle \hat{J}_1 \rangle$ and an increase in $\langle \hat{J}_2 \rangle$, resulting in a uniform rotation about the principal *y* axis, and similarly for $\langle \hat{J}_2 \rangle$. An exception is when the planar rotation is about the $\phi_o$ = 45$^o$ line in the *x-y* plane in which case the planar uniform rotation is stable in CCRM3 but not in MSCRM3, as in the $^{20}$Ne case discussed below.

From the above results, we conclude that CCRM3 predicts only uniform rotation (about the 45$^o$ line in the *x-y* plane or about the *x* or *y* axis with $\theta$=90$^o$), and it does not predict any triaxial rotations. This may or may not be the case in MSCRM3 (as the $^{24}$Mg case discussed below shows).

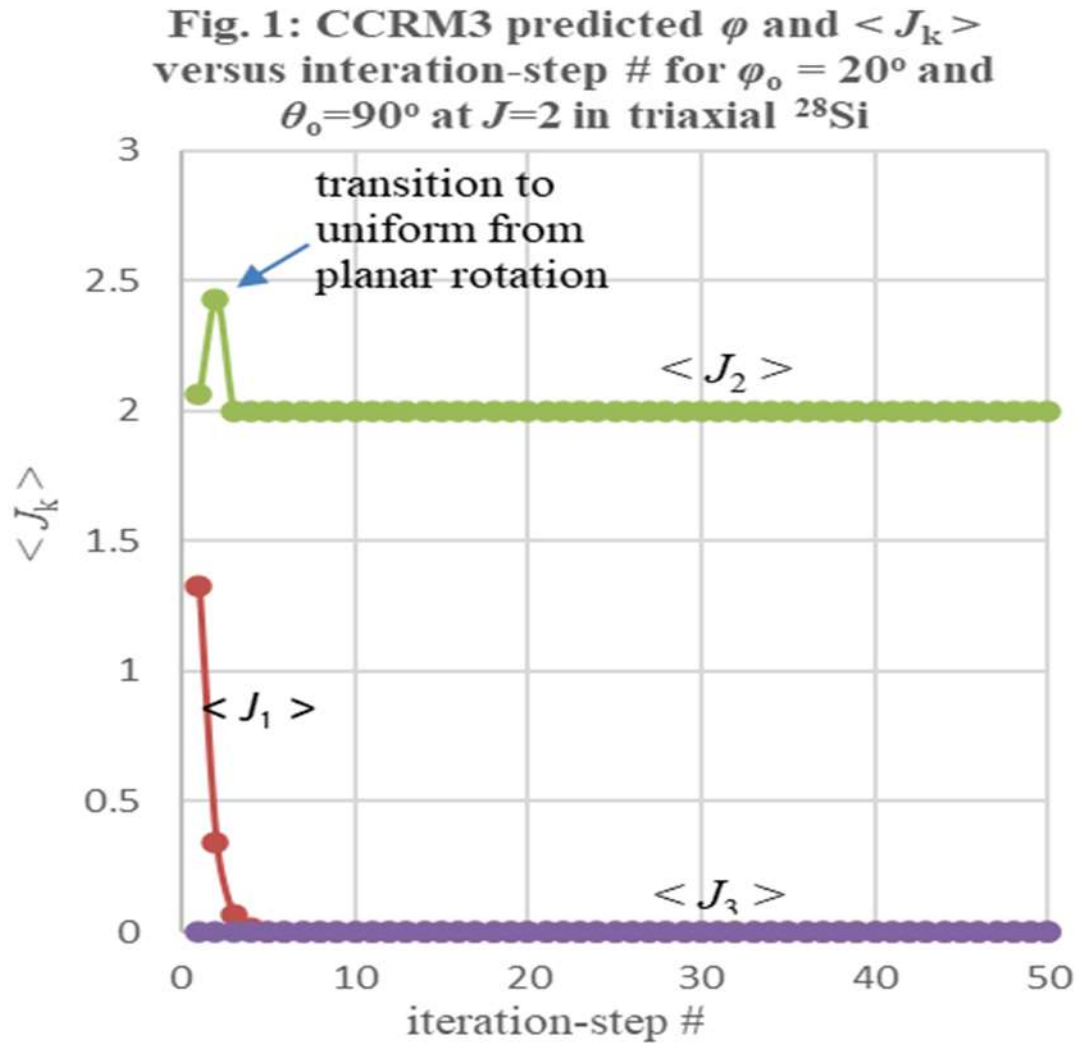

**Fig. 1: CCRM3 predicted $\varphi$ and $< J_k >$ versus interation-step # for $\varphi_o$ = 20° and $\theta_o$=90° at $J$=2 in triaxial $^{28}$Si**

The self-consistency between $\Omega_k$ and $\langle \hat{J}_k \rangle$ in MSCRM3 also allows the intrinsic system to rotationally relax after the imposed $J$ in the first iteration step is removed and replaced by Eq. (28). This rotational relaxation yields a value of $J_{cal}$ that is smaller or larger than $J$. In contrast, CCRM3 forces the intrinsic system to accept the imposed value $J$ that may be inconsistent with the nucleonic motion.

For prolate $^{20}$Ne (axially-symmetric about *z* axis), $\phi_o = 45^o$, and $\theta_o > 0^o$, MSCRM3 predicts a planar rotation in the *x-y* plane at $J_{cal} \leq 6$. At $J_{cal}$ = 7.76, $^{20}$Ne becomes triaxial (due to the interaction between centrifugal stretching and incompressibility condition), and the fluctuations in $\langle \hat{J}_2 \rangle$ are amplified reducing it to zero. This causes a transition to uniform rotation about the *x* axis (i.e., quenching of the wobbly rotation) and the switching from the constraint $\langle \hat{J}^2 \rangle = \hbar^2 J(J+1)$ to $\langle \hat{J}^2 \rangle = \hbar^2 J^2$. At $J_{cal}$ = 8, $^{20}$Ne becomes axially symmetric about the *x* axis and the rotational band terminates. The quenching of the wobbly rotation reduces the energy-level spacing between $J_{cal}$=6 and 8 and improves the agreement with the measured excitation energy as seen in Fig. 2[5]. The remaining discrepancy between the measured and MSCRM3-predicted excitation energies in Fig. 2 is partly due to the neglect of the spin-orbit interaction, which we have estimated to increase the predicted energy by 10%. The still remaining discrepancy is due to the neglect of the residuals of $\hat{J}_k^2$ and $\hat{V}$ in Eq. (16), which is estimated, using MSCRM1 for uniaxial rotation, Tamm-Dancoff approximation, and cranked 1-particle 1-hole basis states, to be 1 to 2 *MeV* increasing the predicted excitation energy close to the measured excitation energy. In contrast, in CCRM3 the uniform planar rotation of $^{20}$Ne about the $\phi_o$=45$^o$ line in the *x-y* plane persists at all *J* values with increasing excitation energy and no rotational-band termination, as seen in Fig. 2.

[4] This result implies that the rotational instability occurs for any self-consistent mean-field potential and not just for the oscillator potential.

[5] This reduction in energy-level spacing has been known [2,4] but has not been attributed to any specific phenomenon. MSCRM3 attributes this phenomenon to the quenching of wobbling at $J_{cal}$ =8.

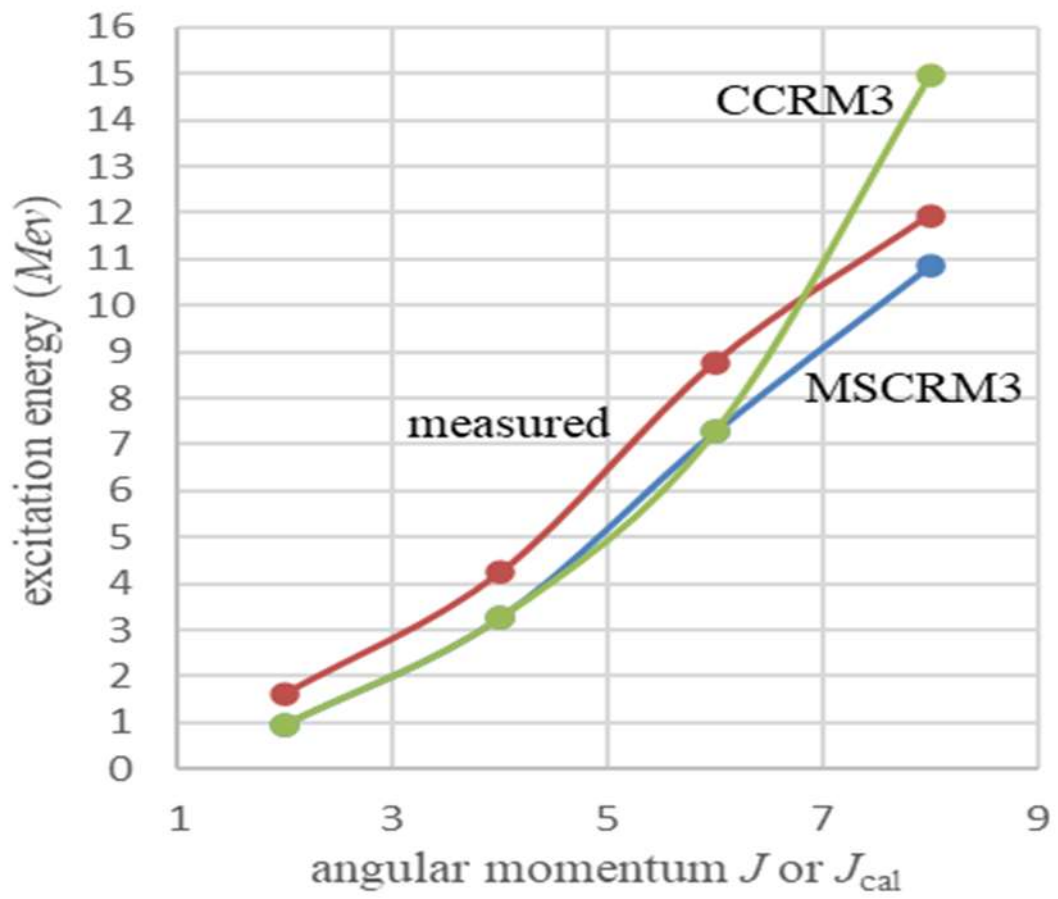

**Fig. 2: Measured and MSCRM3 and CCRM3 predicted excitation energies versus $J$ or $J_{cal}$ for $\varphi_o$ = 45° in prolate $^{20}$Ne**

For prolate $^{24}$Mg and $^{28}$Si, and $\phi_o$=45°, MSCRM3 predicts a wobbly planar rotation in the *x-y* plane at respectively $J_{cal}$<4 and 8. At respectively $J_{cal}$=4 and 10, the rotation transitions to a uniform rotation about the *x* or *y* axis and the nuclei become triaxial. At respectively $J_{cal}$=4 and 12 the nuclei become axially symmetric about the *x* or *y* axis and the rotational band terminates. In contrast, CCRM3 predicts a planar uniform rotation about the 45° line in the *x-y* plane of an axially symmetric (about *z* axis) nucleus at all *J* values and no band termination.

For triaxial $^{24}$Mg, $\phi_o$=90°, and 0°<$\theta_o$<90°, MSCRM3 predicts $\langle \hat{J}_1 \rangle$=0 at all $J_{cal}$ values, and a steady planar rotation in the *y-z* plane at $J_{cal}$ =2 and 4, and a uniform steady rotation about the *y* axis (i.e., $\langle \hat{J}_1 \rangle = \langle \hat{J}_3 \rangle$=0, and $\langle \hat{J}_2 \rangle \neq 0$) at *J*=6. At $J_{cal}$ =8, the nucleus becomes axially symmetric about the *y* axis and the rotational band terminates, in contrast to the case for $\phi_o$ =90° and $\theta_o$=90°, where the nucleus rotates about the *y* axis and remains triaxial at all $J_{cal}$ and the rotational band does not terminate. This result shows that a band termination and its $J_{cal}$ value can be sensitive to the $\theta_o$ value. In contrast, for $\phi_o$=90°, and 0°<$\theta_o \le$ 90°, CCRM3 predicts a uniform rotation about the *y* axis of a triaxial nucleus at all *J* values, and the rotational band does not terminate.

For triaxial $^{24}$Mg, $\phi_o$=89°, and $\theta_o$=54.7° at $J_{cal}$ =2, MSCRM3 predicts that the rotational-state instability reduces $\langle \hat{J}_1 \rangle$ to near zero at the end of the iteration process resulting in a nearly planar triaxial rotation of a triaxial $^{24}$Mg in a plane close to the *y-z* plane as seen in Fig. 3 and 4. Figs. (3) and (4) show respectively the values of the angular-momentum and quadrupole-moment components at each $J_{cal}$ value at the end of the iteration process. A similar triaxial rotation of triaxial $^{24}$Mg is predicted at $J_{cal}$=4. At *J*=6, the rotation becomes uniform along the *y* axis and $^{24}$Mg remains triaxial. At $J_{cal}$ =8, $^{24}$Mg remains triaxial and the uniform rotation transitions to a triaxial rotation, as seen in Figs. 3 and 4. At $J_{cal}$=10 the nuclear rotation and shape become almost axially symmetric about the *x* axis, as seen in Figs. 3 and 4. At $J_{cal}$ *J*=12, the nuclear rotation and shape become spherically symmetric and the rotational band terminates.

**Fig. 3: MSCRM3 predicted < $J_k$ > versus $J_{cal}$ at end of iteration process for $\varphi_o$=89° and $\theta_o$=54.7° in triaxial $^{24}$Mg**

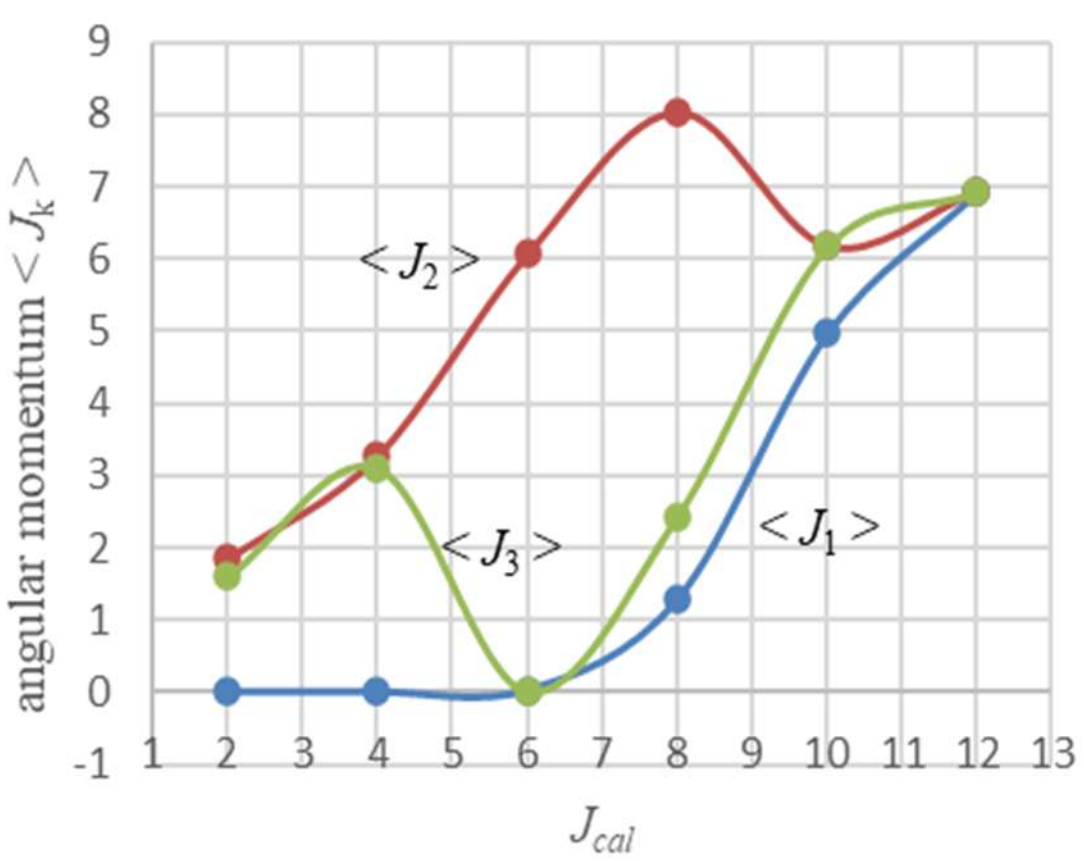

**Fig. 4: MSCRM3 predicted quadrupole moments < $x_k^2$ > versus $J_{cal}$ for $\varphi_o$=89° and $\theta_o$=54.7° in triaxial $^{24}$Mg**

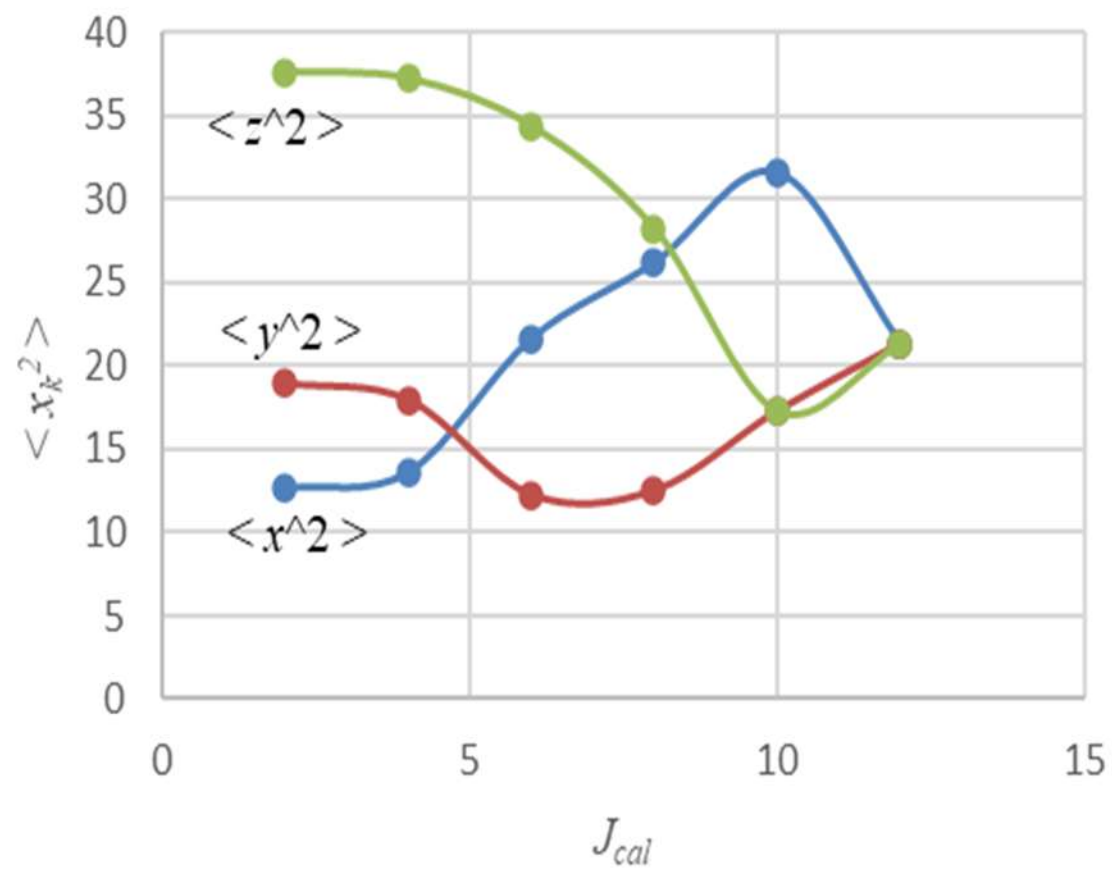

For triaxial $^{24}$Mg, $\phi_o$=89° and $\theta_o$=90° at $J_{cal}$=2, MSCRM3 predicts, during the iteration process, a steady rotation about the *y* axis with very small values of $\langle \hat{J}_1 \rangle$ and $\langle \hat{J}_3 \rangle$, which begin to slowly increase to $J_{\text{cal}}$=13 and 10 respectively because of competing effects of the three terms in the first of Eq. (24). At $J_{\text{cal}}$=12, $\langle \hat{J}_1 \rangle$ and $\langle \hat{J}_3 \rangle$ become comparable to $\langle \hat{J}_2 \rangle$. Therefore, the uniform rotation about the *y* axis transitions to a triaxial rotation of a triaxial nucleus and eventually to a spherically symmetric rotation of a spherical nucleus, at which point the rotational band terminates. This result contrasts with that for the case $\phi_o$=89°, and $\theta_o$=54.7° described above. In contrast, CCRM3 predicts a uniform rotation of triaxial nucleus about the *x* axis at *J* =2,4,6, and 8. At *J*=8, the nucleus becomes axially symmetric about the *x* axis and the rotational band terminates.

For triaxial $^{24}$Mg, 21°≤$\phi_o$<89°, and $\theta_o$=90°, at 2≤$J_{\text{cal}}$<12, MSCRM3 predicts a steady uniform rotation about the *y* axis of a triaxial nucleus. At $J_{\text{cal}}$ =12, the nucleus and the rotation become spherically symmetric and the rotational band terminates, similarly to that for $\phi_o$=89° and $\theta_o$=90° at $J_{\text{cal}}$ =12.

For triaxial $^{24}$Mg, 0°<$\phi_o$≤20°, and $\theta_o$=90°, MSCRM3 predicts no rotation of any kind at any $J_{\text{cal}}$; for $\phi_o$=0° and $\theta_o$=90°, MSCRM3 predicts a uniform rotation about the *x* axis at 2≤ $J_{\text{cal}}$ <8. At $J_{\text{cal}}$ =8, the nucleus becomes axially symmetric about the *x* axis and the rotational band terminates. In contrast, for 0°≤ $\phi_o$<90°, and 0°<$\theta_o$≤90°, CCRM3 predicts a uniform rotation of triaxial nucleus about the *x* axis at *J*=2,4,6, and 8. At *J*=8, the nucleus becomes axially symmetric about the *x* axis and the rotational band terminates; and for $\phi_o$=90° and 0<$\theta_o$≤90°, CCRM3 predicts a uniform rotation about the *y* axis of a triaxial nucleus at all *J* values, and the rotational band does not terminate.

Shell-model calculation [30], experiments [31], and MSCRM3 prediction for 21°≤$\phi_o$≤89° and 0°< $\theta_o$≤90° presented above show that the rotational band in $^{24}$Mg terminates at *J*=12 when the nucleus and rotation about the *y* axis become spherically symmetric. In contrast, CCRM3 predicts that the rotational band terminates at *J*=8 when the nucleus becomes axially symmetric about the *x* axis.

The above results also show that CCRM3 predicts only uniform principal-axis rotation of a triaxial $^{24}$Mg. This result may not be in conflict with the conclusions from the CCRM3 analysis in [19] because [19] used only one kind of nucleons.

For triaxial $^{28}$Si, neither MSCRM3 nor CCRM3 predict any physically meaningful collective rotation for any *J*, $J_{\text{cal}}$, $\phi_o$, and $\theta_o$ values.

For oblate $^{24}$Mg and $^{28}$Si, MSCRM3 and CCRM3 predict no steady stable collective rotation of any kind.

## 4. Summary and conclusions

The conventional cranking model for uniaxial rotation is frequently used in nuclear structure studies. The model is semi-classical, phenomenological, and uses a constant adjustable angular-velocity parameter. It is, therefore, time-reversal and $D_2$ non-invariant. To investigate the impact of a microscopic (dynamic) angular velocity on the results of such studies, a quantum microscopic self-consistent, cranking model Schrodinger equation for triaxial rotation (MSCRM3) is derived in two steps.

In the first step, we apply a dynamic rotationally-invariant exponential rotation operator to a rotationally-invariant nuclear Schrodinger equation to transform it to that of a rotationally and time-reversal invariant rotor Schrodinger equation. The rotor Hamiltonian is made optimally intrinsic by choosing the three rotation angles in the rotation operator to be given by a collective rigid-flow velocity-field prescription and to be canonically conjugate to the angular momentum.

In the second step, Hartree-Fock variation is applied to the rotor Schrodinger equation to obtain MSCRM3 Schrodinger equation with a HF mean-field Hamiltonian $\hat{H}_{crHF}$ plus residuals of the square of the angular-momentum operator and two-body interaction and other negligibly small correction terms. The MSCRM3 Schrodinger equation is identical in form to that of the conventional cranking-model equation for triaxial rotation (CCRM3) except that in MSCRM3 the angular velocity is not a constant parameter but is a dynamical quantity given by the ratio of the expectations of the angular momentum and rigid-flow kinematic moment of inertia. As a consequence, $\hat{H}_{crHF}$ is time-reversal and $D_2$ invariant, rendering the MSCRM3 wavefunction a superposition of either even or odd angular-momentum eigenstates, in contrast to the CCRM3 wavefunction, which is a superposition of even and odd angular momentum eigenstates.

For the self-consistent deformed harmonic oscillator (mean-field) potential, the MSCRM3 and CCRM3 governing equations (or equivalently the three normal-mode frequencies of $\hat{H}_{crHF}$ ), coupled to those of the density-potential self-consistency and constant-volume condition, are solved iteratively. The simple deformed harmonic-oscillator potential energy is used to facilitate the comparison of the MSCRM3 predicted results with those of CCRM3 and with those of the earlier CCRM3 predictions that used this potential. The objective of the analysis in this article is to determine the differences between the predictions of the two models in a transparent way with the fewest number of adjustable parameters, and not to compare them with measured data, except for $^{20}$Ne case discussed below.

A summary of the major differences between the predictions of MSCRM3 and CCRM3 for prolate $^{20}$Ne, $^{24}$Mg, and $^{28}$Si, and oblate and triaxial $^{24}$Mg, and $^{28}$Si is given below.

CCRM3 predicts only uniform rotations whereas MSCRM3 generally predicts uniform, wobbly, planar, and triaxial rotations.

CCRM3 and MSCRM3 predict that the coupling between the intrinsic and rotational motions generated by the density-potential self-consistency-condition renders the rotational states unstable, causing them to decay to zero angular momentum unless prevented by an angular-momentum constraint on the wavefunctions.

In CCRM3, the rotational-state instability causes the initial triaxial rotation to transition to a planar uniform rotation in the *x*-*y* plane and subsequently to a uniform rotation about the *x* or *y* axis with a rotational-band termination at *J*=8,4, and 8 in prolate $^{20}$Ne, $^{24}$Mg, and $^{28}$Si respectively. An exception to this situation is when the initial rotation is axially symmetric, in which case it remains so at all *J* values with no band termination. Therefore, CCRM3 does not predict any steady stable triaxial rotation in any of the nuclei $^{20}$Ne, $^{24}$Mg, and $^{28}$Si. These results may not be in conflict with the CCRM3 analysis result in [19] because [19] used only one kind of nucleons.

MSCRM3 predicts uniform principal-axis rotation and planar non-uniform (i.e., wobbly) rotation in the *x*-*y* plane and band termination at *J*=8 and 4, and 12 for prolate $^{20}$Ne, $^{24}$Mg, and $^{28}$Si respectively (and also at *J*=8 for initially non-uniform rotation in prolate $^{28}$Si). For planar wobbly initially axially-symmetric (i.e., along the 45° line in the *x*-*y* plane) rotation in $^{20}$Ne, MSCRM3 predicts that at *J*=8 the rotation transitions to a uniform rotation about the *x* axis. This transition (i.e., quenching of the wobbly rotation) reduces the energy-level spacing between *J*=6 and 8 as is observed in experiments [2,4]. This result resolves the long-standing mystery as to the cause of this energy-level spacing reduction, which is not predicted by CCRM3. It is shown that the residuals of the square of the angular momentum and two-body interaction, and to a lesser extent the spin-orbit interaction in MSCRM3 remove the remaining discrepancy between the predicted and measured excitation energies in $^{20}$Ne.

For triaxial $^{24}$Mg, MSCRM3 and CCRM3 predict band termination at *J*=4 and 8 (depending on initial angular-velocity polar angles) when the nucleus becomes axially symmetric. However, MSCRM3 also predicts band termination at *J*=12 (observed in shell-model calculations [30] and experiments [31]) when $^{24}$Mg and its rotation become spherically symmetric about one of the principal axes. This latter band termination is not predicted by CCRM3.

For some values of the angular-velocity polar angles at the start of the iteration process, MSCRM3 predicts no band termination in triaxial $^{24}$Mg whereas CCRM3 does, and vice versa.

In a number of cases for triaxial $^{24}$Mg, MSCRM3 predicts no rotation of any kind over a range of angular-velocity polar angles whereas CCRM3 predicts uniform rotation.

In a number of cases MSCRM3 predicts band termination whereas CCRM3 does not and vice versa.

It is shown that neither MSCRM3 nor CCRM3 predict any physically meaningful rotations in triaxial $^{28}$Si for any values of *J* and the initial angular-velocity polar angles.

No steady stable collective rotation of any kind is predicted by neither MSCRM3 nor CCRM3 for oblate $^{24}$Mg and $^{28}$Si.

The above results show that CCRM3 does not predict any of the three-dimensional phenomena predicted by MSCRM3 such as reduced energy level spacing in $^{20}$Ne and band terminations in prolate $^{28}$Si and triaxial $^{24}$Mg at *J*=12 when the nuclei and their rotations become spherically symmetric. These three-dimensional phenomena are induced by the action of the microscopic (dynamic) angular velocity in MSCRM3. We, therefore, conclude that CCRM3 is effectively a uniaxial rotation model. Therefore, using CCRM3 or its uniaxial version, one would miss capturing three-dimensional rotation phenomena.

In future, it may be useful and instructive to study the differences between the two model predictions for the above and other nuclei and for more realistic interactions. It may also be useful to generalize MSCRM3 to include vibrational excitations.

## Acknowledgements

The author would like to thank Professor M. A. Caprio at the University of Notre Dame for his encouragements and providing valuable comments on the manuscript.